\documentclass[%
 reprint,
nofootinbib,
 amsmath,amssymb,
 aps,
]{revtex4-1}

\usepackage{graphicx}
\usepackage{dcolumn}
\usepackage{bm}
\usepackage{multirow}
\usepackage{amssymb,mathrsfs,amsmath}
\usepackage{amsfonts}
\usepackage{color,xcolor}
\usepackage{subfigure}
\def\linkcolor{cyan!70!black}

\usepackage[
colorlinks=true
,urlcolor=\linkcolor
,anchorcolor=\linkcolor
,citecolor=\linkcolor
,filecolor=\linkcolor
,linkcolor=\linkcolor
,menucolor=\linkcolor
,linktocpage=true
,pdfproducer=medialab
,pdfa=true
]{hyperref}
\usepackage{orcidlink}
\usepackage{soul}

\newcommand\ptwiddle[1]{\mathord{\mathop{#1}\limits^{\scriptscriptstyle(\sim)}}}

\begin{document}
\preprint{HEP-IFUNAM, 24-1}

\title{Exploring hadronic de-excitation via Lepton Flavor Violation}

\author{Fabiola Fortuna\,\orcidlink{0000-0002-8938-7613}}%
\email{fabyfortuna@fisica.unam.mx}
\affiliation{Instituto de F\'{\i}sica,  Universidad Nacional Aut\'onoma de M\'exico, AP 20-364,  Ciudad de México 01000, M\'exico}
\author{L. Esparza-Arellano\,\orcidlink{0009-0007-9282-6131}}%
\email{leonardoesparza@estudiantes.fisica.unam.mx}
\affiliation{Instituto de F\'{\i}sica,  Universidad Nacional Aut\'onoma de M\'exico, AP 20-364,  Ciudad de México 01000, M\'exico}%
\author{Genaro Toledo\,\orcidlink{0000-0003-4824-9983}}%
 \email{toledo@fisica.unam.mx}
\affiliation{Instituto de F\'{\i}sica,  Universidad Nacional Aut\'onoma de M\'exico, AP 20-364,  Ciudad de México 01000, M\'exico}%

\date{\today}

\begin{abstract}
In this work, we consider a hadronic de-excitation mechanism via lepton flavor violation (LFV). Namely, for vector and pseudoscalar meson decays of the form $V^\prime \to V \mu e$ and  $P^\prime \to P \mu e$, respectively. They are described considering effective dim-5 and dim-7 operators. Using the current bounds for the effective couplings obtained from direct processes, we exhibit the different features in the dilepton invariant mass distribution and the branching ratios, depending on the effective operator. 
We consider hadronic states, from light mesons to quarkonia. We show that the branching ratios of quarkonia system  for LFV are comparable with the estimates in other hadronic systems.
Our results show that the dilepton invariant mass may be useful to constrain individual contributions and help disentangle them, when complemented with observables from nuclei.

\end{abstract}

\maketitle

\section{Introduction}
 
The search for signals of LFV has been performed in a wide range of scenarios. 
Those involving direct processes, such as $\mu \to e \gamma$, have a current experimental bound on the branching ratio (BR) of $\mathcal{O}(10^{-13})$ \cite{MEGII:2023ltw}. They offer clean bounds in the sense that they are purely leptonic processes and are expected to be further improved in the near future. 
MEG II expects to surpass its current bound on $\mu \to e \gamma$ to reach a sensitivity of $6 \times 10^{-14}$ \cite{MEGII:2018kmf};
SINDRUM II has measured the muon to electron conversion in muonic gold with a precision of $7 \times 10^{-13}$ \cite{SINDRUMII:2006dvw};
 MU3e is expect to measure the $\mu \to eee$ decay with a branching ratio of $\mathcal{O}(10^{-16})$  \cite{Mu3e:2020gyw}; COMET experiment will measure the $\mu \to e$ conversion in aluminum nuclei with a precision, in phase I, of $\mathcal{O}(10^{-15})$ \cite{COMET:2018auw}; DeeME will get a conversion rate in carbon of $\mathcal{O}(10^{-13})$ \cite{Natori:2014yba} and the Mu2e experiment, in phase I, will reach a sensitivity of $\mathcal{O}(10^{-16})$ on aluminum \cite{Mu2e:2022ggl}. LFV is also a topic of interest for US-led initiatives on Rare Processes and Precision Measurements Frontier \cite{Davidson:2022jai}, which can test new physics scales of thousands of TeV.

LFV processes can be described in general grounds, considering EFT or particular models responsible for the transition \cite{Kuno:1999jp,deGouvea:2013zba}.
In the minimal extension of the SM the $\mu \to e \gamma$ branching ratio is set to $\mathcal{O}(10^{-54})$ \cite{Petcov:1976ff, Bilenky:1977du}, leaving a wide parameter space to look for non-standard contributions \cite{Fernandez-Martinez:2024bxg}. Possible intermediate states consider heavy Majorana neutrinos, where the bounds for masses and couplings are placed from direct or indirect information \cite{Abada:2015zea}. 
The search of signatures of non-standard decays in other processes, such as in hyperon decays, can lead to very restrictive indirect bounds \cite{Hernandez-Tome:2021byt}.
 LFV in meson decays have been explored considering quark transitions of the form $q \to q^\prime l^+_i l^-_j $\cite{Glashow:2014iga,Calibbi:2017uvl,Becirevic:2016zri,Descotes-Genon:2023pen}, where they necessarily involve different quark structure of the initial and final mesons. The use of effective operators, including both lepton and quark fields, provides useful information to place bounds on the branching ratio of such decays. These transitions have been studied for heavy to light mesons within the effective field theory (EFT) framework, for instance, in Ref.~\cite{Descotes-Genon:2023pen,Plakias:2023esq, Ali:2023kua}, where LHC data was used. They derived upper limits on LFV decays of the $B$ meson into another meson $M$, of the form $B \to M l_i^+l_j^-$.\\
LFV vector meson annihilation into two leptons, in the minimal extension of the SM, would lead to a BR smaller than $10^{-50}$ \cite{Abada:2015zea}.
Studies in EFT approaches, have obtained strong bounds for the BR's of quarkonia \cite{Calibbi:2022ddo}. Indirect bounds for the BR, depending on the corresponding operator, can be of $\mathcal{O} (10^{-25})$ for the $\Upsilon(2S) \to \mu e$ and $\mathcal{O}(10^{-27})$ for the $\psi(2S) \to \mu e$ \cite{Calibbi:2022ddo}. In Table \ref{tab:others}, we show the bounds for the BR's, obtained for LFV annihilation of ground and excited vector mesons, considering particular model realizations \cite{Abada:2015zea,Dong:2024lnk,Yue:2016mqm,Sun:2022zbq}. These processes will serve as a reference for the ones of interest in this study. On the experimental side, Table~\ref{tab:exp} presents the current upper limits on LFV decays of neutral vector mesons into a muon–electron pair.

\begin{table}[htb]
\begin{center}
\renewcommand{\arraystretch}{1.5}

\begin{tabular}{|l|c|c|}
\hline
   & BR($J/\psi\to \mu e$) & BR($\Upsilon(1S)\to \mu e$) \\
\hline
Ref.~\cite{Abada:2015zea} EFT model & $<3\times 10^{-24}$ & $<2\times 10^{-24}$\\
Ref.\cite{Abada:2015zea} ISS    & $<9\times 10^{-20}$ & $<2\times 10^{-17}$\\
Ref.~\cite{Yue:2016mqm} new $Z^\prime$ & $3.9\times 10^{-19}$ & $<1.8\times 10^{-16}$\\
Ref.~\cite{Sun:2022zbq} MRSSM & $\sim 10^{-24}$ & $\sim 10^{-22}$\\
\hline
   & BR($\psi(2S)\to \mu e$) & BR($\Upsilon(2S)\to \mu e$) \\
\hline
Ref.~\cite{Yue:2016mqm} new $Z^\prime$ & $1\times 10^{-19}$ & $<2.0\times 10^{-16}$\\
Ref.~\cite{Sun:2022zbq} MRSSM & $\sim 10^{-25}$ & $\sim 10^{-22}$\\
\hline
\end{tabular}
\caption{Overview of bounds for the BR's reported in the literature, of ground and excited vector mesons LFV annihilation processes, considering particular model realizations. ISS and MRSSM stand for inverse seesaw mechanism and minimal R-symmetric supersymmetric standard model, respectively.} \label{tab:others} 
\end{center}
\end{table}

\begin{table}[htb]
\begin{center}
\renewcommand{\arraystretch}{1.5}
\begin{tabular}{|c|c|c|}
\hline
 Decay  & Limit & Experiment \\
\hline
$J/\psi\to \mu e$ & $4.5\times 10^{-9}$ & BESIII (2022) \cite{BESIII:2020nme} \\
$\Upsilon(1S)\to\mu e$ & $3.9\times 10^{-7}$ & Belle (2022) \cite{Belle:2022cce}\\
$\Upsilon(2S)\to \mu e$ & $3.6\times 10^{-7}$ & BABAR(2022) \cite{BaBar:2021loj}\\
\hline
\end{tabular}
\caption{Current experimental bounds for the BR's of vector quarkonium LFV annihilation into a $\mu e$ pair.} \label{tab:exp}  
\end{center}
\end{table}

A scenario that has received less attention is at the interplay of those mentioned above. Namely, the one involving hadronic states with the same initial and final quarks. This feature has been used, at the nuclear level, to place bounds on the LFV in nuclei, and has proven to be a possible way to deepen the understanding of LFV in nuclear matter, in experiments such as NA64~\cite{NA64}. In this case, the limitations are set by the nuclear effects to be properly accounted for at the given precision.
Using an EFT approach to describe such processes \cite{Fortuna:2023paj} allows to explore dim-5 and dim-7 operators involving one and two photons intermediate states, which then produce the LFV pair. The dim-5 and dim-7 effective operators have been previously used to explore LFV leptonic decays and LFV in nuclei 
\cite{Davidson:2018kud,Davidson:2020ord,Fortuna:2022sxt, Fortuna:2023paj,Cirigliano:2022ekw}.

A possible hadronic scenario for the same effective approach, which has not been considered in the literature, is the use of radially excited hadronic states, whose de-excitation process goes via the LFV mechanism. One example is the excited vector meson ($V^\prime$) which decays into a ground state meson ($V$) and the LFV pair, as depicted in Figs. \ref{fig:dim5} and \ref{fig:dim7}. 
A similar mechanism can be applied for Pseudoscalar excited mesons ($P^\prime$) which decays into a ground state meson ($P$), as depicted in Fig. \ref{fig:dim7_p}. These scenarios are cleaner channels to look for LFV than the conversion processes in nuclei, as they are free from nuclear effects, although proper knowledge of the hadronic form factors of the transitions and larger statistics are required. Particular realizations can be found for light and heavy mesons. In the light mesons sector a typical example is the radially excited vector meson $\rho (1450)$, generally denoted as the $\rho^\prime$ meson, which is observed in low energy experiments with a large decay width \cite{Diekmann:1986wq,Kurdadze:1983ys, Achasov:2016zvn,BABAR:2021cde,BESIII:2019gjz,Belle:2008xpe}.
For bottomonia, the relatively small decay width of the meson (compared to the light ones) and the expected large statistics from experiments, such as LHCb \cite{LHCb:2019ujz} and Belle II \cite{Belle:2023jwr,HernandezVillanueva:2022dpt}, might offer the possibility of placing bounds on these unexplored processes. For example, $158\times10^{6}$  $\Upsilon(2S)$ events have been already produced in Belle \cite{Belle:2022cce}, and a large increase is expected for Belle II. Similar features are foreseen for charmonia, where experiments such as BES III are also making progress in this direction \cite{BESIII:2020nme,BESIII:2021slj}. Future experiments as the STCF could also produce a large number of charmonia \cite{Lyu:2021tlb}.

In this work, we illustrate the main features of the hadronic de-excitation involving LFV. We consider particular systems for light, intermediate and heavy mesons (quarkonia). We explore the role of the effective operators involved, determining the branching ratio and the di-lepton invariant mass distribution, consistent with the current bounds for the effective couplings. We analyze to what extent they can offer new features of LFV from hadronic states. We use the EFT approach to outline the LFV vertices \cite{Fortuna:2022sxt}. The hadronic side is described using the general structure for the Vector-Pseudoscalar-Photon interaction. The corresponding couplings are determined using the vector meson dominance model (VMD) \cite{Bando:1984ej,Fujiwara:1984mp,Meissner:1987ge,Avalos:2022txl} for the light  mesons and quark models \cite{Ridwan:2024ngc,Li:2018uif} for quarkonia.
In Section II, we introduce the dim-5 and dim-7 effective operators that describe the charged LFV processes, and the constraints that we derive on the effective coefficients from a direct experimental search. 
In Section III we outline the amplitude for the decay processes using the corresponding dim-5 or dim-7 effective operator and their hadronic transitions. 
Dim-5 effective operators apply exclusively to the
$V^\prime\to V\mu e$ decay at tree level, while dim-7 effective operators apply to both $V^\prime\to V\mu e$ and $P^\prime\to P\mu e$ decays at one-loop level. 
Section IV is devoted to present our results on the dilepton invariant mass and branching ratio, for all the systems under consideration. Finally, we discuss the different features associated with each operator and present our conclusions. 


\section{Effective field theory description}

In this work we use effective operators that generate charged lepton flavor violating interactions.
Particularly, we focus on the $\mu e \gamma$ and $\mu e \gamma\gamma$ effective interactions. The former is generated by the dim-5 dipole operator 
\begin{equation}\label{eq:dim5}
\mathcal{L}_\text{dim-5} = 
D_{R}^{\mu e}\, \bar{\mu}_{L} \sigma_{\mu\nu} e_{R} F^{\mu\nu} + D _{L}^{\mu e}\, \bar{\mu}_{R} \sigma_{\mu\nu} e_{L} F^{\mu\nu} +{ \rm h.c.}\,,
\end{equation}
while the latter is produced by the dim-7 diphoton operator
\begin{align} \label{eq:dim7}
  \mathcal{L}_\text{dim-7}&=\left(G_{SR}^{\, \mu e}\bar{\mu}_{L}e_{R}+G_{SL}^{\, \mu e}\bar{\mu}_{R}e_{L}\right)F_{\mu\nu} F^{\mu\nu}\nonumber\\ 
&+\left(\tilde{G}_{SR}^{\,\mu e} \bar{\mu}_{L}e_{R}+\tilde{G}_{SL}^{\,\mu e} \bar{\mu}_{R}e_{L}\right)\tilde{F}_{\mu\nu}F^{\mu\nu} +
{\rm h. c.}\,,
\end{align}
where $\tilde{F}_{\mu\nu}=\frac{1}{2}\epsilon_{\mu\nu\sigma\lambda}F^{\sigma\lambda}$ is the dual electromagnetic field-strength tensor ~\cite{Bowman:1978kz}.
The subscripts $L(R)$ in the equations above indicate the chirality of the lepton.

Recently, in Ref. \cite{Fortuna:2022sxt} these operators have been used to obtain indirect upper limits in the $\ell_i\to\ell_j\gamma\gamma$ decays from the experimental limits on the $\ell_i\to\ell_j\gamma$ processes. Taking the decay rates as follows: 
\begin{align}\label{eq:rate_gamma}
 \Gamma(\mu \to e \gamma)\big|_{\rm dim-5}&= \frac{m_\mu^3}{4 \pi }\big|D^{\mu e}\big|^2 \,,\nonumber\\
 \Gamma(\mu \to e \gamma)\big|_{\rm dim-7}&\sim \frac{\alpha \big|G_{\mu e}\big|^2}{256\pi^4}m_{\mu}^7\log^2\left(\frac{\Lambda^2}{m_\mu^2}\right)\,,
\end{align}
where $\big|D^{\mu e}\big|^2=\big|D_R^{\mu e}\big|^2+\big|D_L^{\mu e}\big|^2$ and $\big|G_{\mu e}\big|^2=\big|G_{SR}^{\mu e}\big|^2+\big|G_{SL}^{\mu e}\big|^2+\big|\tilde{G}_{SR}^{\mu e}\big|^2+\big|\tilde{G}_{SL}^{\mu e}\big|^2$,
and comparing them with the current experimental upper limit on the $\mu\to e\gamma$ decay~\cite{MEGII:2023ltw} we updated the constraints in the $D^{\mu e}$ and $G_{\mu e}$ effective coefficients. These constraints are shown in Table \ref{tab:constraints} .
In the decay rate computed with the dim-7 operators, we are taking only the leading term, and use $\Lambda=100$ GeV to derive the constraint on $G_{\mu e}$ shown in Table \ref{tab:constraints}.

\begin{table}[htb]
\begin{center}
\renewcommand{\arraystretch}{1.5}
\begin{tabular}{|c|c|}
\hline
Coefficient   & Constraint \\
\hline
$\big|D^{\mu e}\big|$    & $3.1\times 10^{-14}$ GeV$^{-1}$\\
$\big|G_{\mu e}\big|$    & $1.1\times 10^{-10}$ GeV$^{-3}$\\
\hline
\end{tabular}
\caption{Constraints on the effective coefficients $D^{\mu e}$ and $G_{\mu e}$ from the current experimental upper limit  on $\mu\to e\gamma$ decay (BR($\mu\to e\gamma) < 3.1\times10^{-13}$  at 90\%CL)~\cite{MEGII:2023ltw}.} \label{tab:constraints}  
\end{center}
\end{table}

Due to the expected sensitivity for $\mu\to e$ conversion process of $\mathcal{O}(10^{-16})$, according to upcoming experiments~\cite{Mu2e:2022ggl}, we expect that in the future the best probe for $\mu e\gamma$ and $\mu e\gamma\gamma$ operators will come from this conversion process. New-physics scenarios where the inelastic $\mu\to e$ conversion can provide additional information on LFV are explored in Ref.~\cite{Haxton:2024ecp}.

\section{LFV de-excitation decay amplitudes}
We will consider particular meson realizations of the de-excitation. For definiteness, for light vector mesons we use the $\rho (1450)$, for charmonium, we use $\psi(2S)$ and for bottomonium $\Upsilon(2S)$. In the case of pseudoscalar mesons we use $\eta_c(2S)$ and $\eta_b(2S)$ for  charmonium and bottomonium, respectively.

In the following we describe the amplitude considering 
dim-5 effective operators, which applies exclusively to the
$V^\prime\to V\mu e$ decay.
Then, we compute the amplitudes considering dim-7 effective operators for both $V^\prime\to V\mu e$ and $P^\prime\to P\mu e$ decays. The hadronic transitions are described in general grounds, using a vector meson dominance approach.\\

\begin{figure}[t]
\includegraphics[scale=0.48]{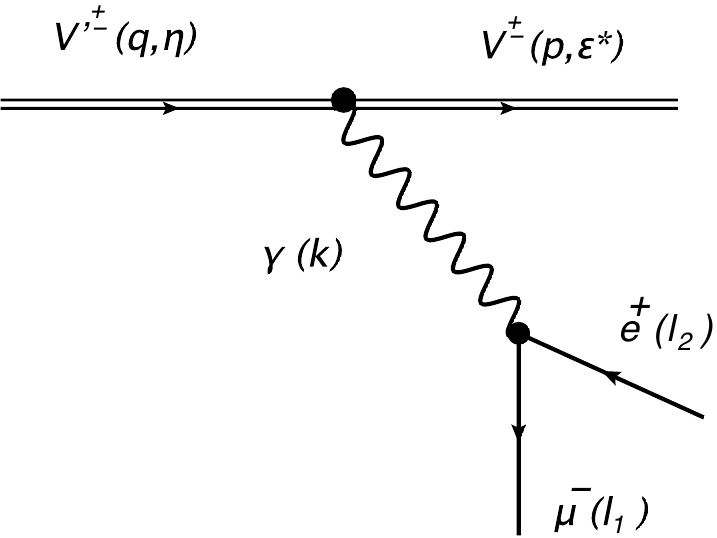}
\caption{\label{fig:dim5} $V^\prime \to V \mu e$ decay driven by the dim-5 effective operators.}
\end{figure}

\begin{figure}[t]
\includegraphics[scale=0.48]{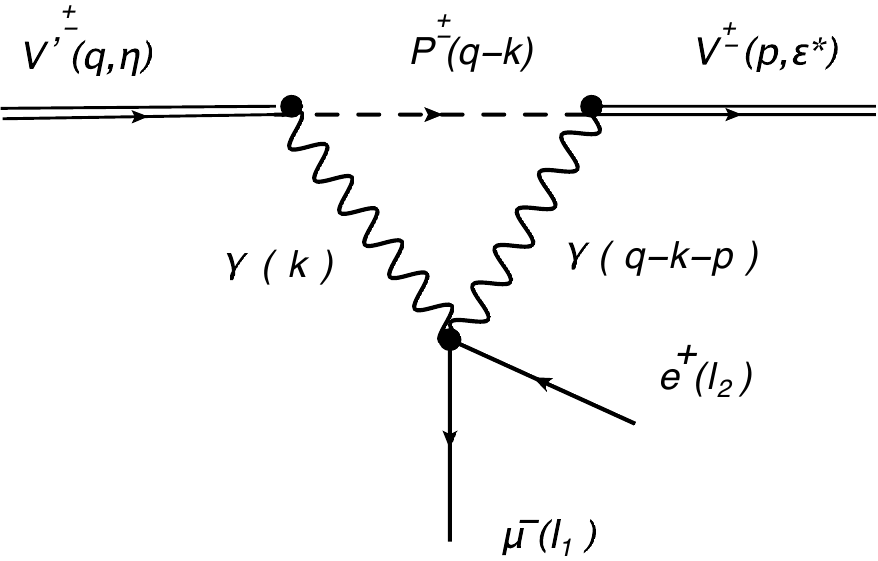}
\caption{\label{fig:dim7} $V^\prime \to V \mu e$ decay driven by the dim-7 effective operators.} 
\end{figure}

\begin{figure}[t]
\includegraphics[scale=0.48]{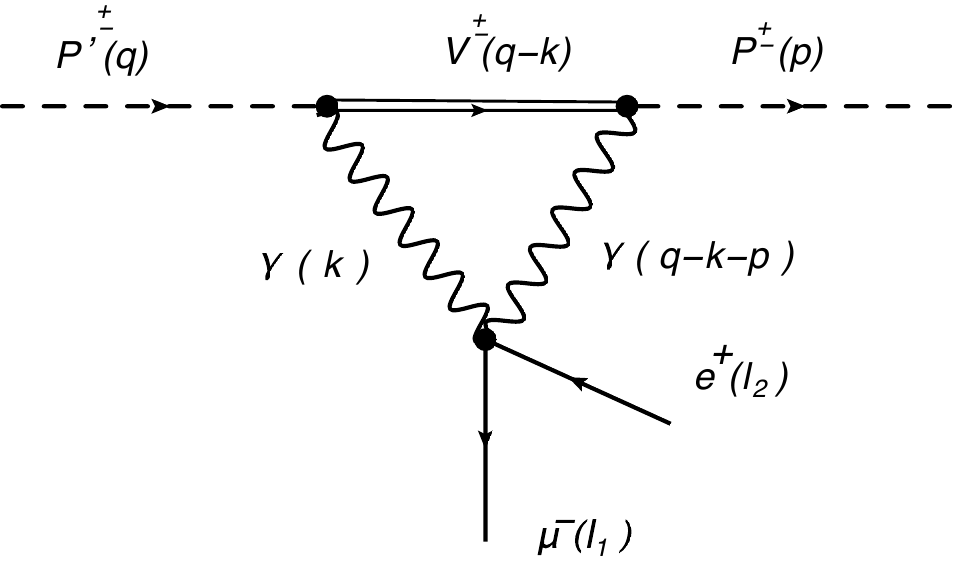}
\caption{\label{fig:dim7_p} $P^\prime \to P \mu e$ decay driven by the dim-7 effective operators.} 
\end{figure}

\subsection{Dim-5 operator driven process}
Let us set the notation for the process as
$V^{\prime \pm}(q,\eta) \to V^{\pm}(p,\epsilon^*) \mu^-(l_1) e^+(l_2)$ decay, where $\eta$ and $\epsilon^*$ correspond to the polarization vectors of the $V^\prime$ and $V$ mesons, respectively.

We first calculate the decay when the dominant contribution is driven by the dim-5 operators in Eq. (\ref{eq:dim5}). This allows to describe the process at tree level, as shown in Fig. \ref{fig:dim5}. The amplitude for this diagram can be written as:
\begin{equation} \label{eq:amp_dim5}
    \mathcal{M}_{\rm dim5} =-\frac{e\, g_{V^{\prime} V \gamma}}{k^2}  \ell_{\mu\nu}(k^\mu g^{\nu\tau}-k^\nu g^{\mu\tau}) \Gamma_{\delta\theta\tau}(q,k) \eta^\delta \epsilon^{*\theta}\,,
\end{equation}
where $k=q-p$ is the transferred momentum into the lepton pair, described by $\ell_{\mu\nu}= \bar{u}_1\sigma_{\mu\nu} (D_{R}^{\mu e} P_R + D_{L}^{\mu e}\, P_L)v_2$, according to the effective operator in Eq. (\ref{eq:dim5}).  On the hadronic side, the $V^\prime \to V \gamma$ transition is modeled following the general structure expected for a $V'V\gamma$ vertex, with two distinct vectors ($V$ and $V'$) and a photon~\cite{Nieves:1996ff, Hagiwara:1986vm}. Thus, the corresponding vertex, $\Gamma_{\delta\theta\tau}(q,k)$, can be parameterized in terms of the magnetic dipole moment and electric quadrupole transition operators (the electric charge form factor is null), as follows:

\begin{align} \label{eq:dip-cuad}
    &\Gamma^{\delta \theta \tau}(q,k)=\beta(g^{\delta\theta}k^\tau-g^{\tau\delta}k^\theta)\nonumber\\
    &\hspace{3mm}+\frac{\gamma}{2m_{V^\prime}^2}\left[ (2q-k)^\tau k^\delta k^\theta-q\cdot k(g^{\theta\tau}k^\delta+g^{\tau\delta}k^\theta) \right].
\end{align}
Notice that each contribution is gauge invariant by itself. The corresponding strengths $\beta$ and $\gamma$, for the dipole and quadrupole interactions, respectively, are free parameters. We are not aware of a quark model estimate of such couplings for the particular cases we are considering \footnote{The radiative decay of radial excitations considered in quark models are typically of the form  $V \to P \gamma$ \cite{Iachello:1992yu,Godfrey:1985xj}.}. However, in transitions of the form $VV\gamma$ they are expected to be around $\beta=2$ and $\gamma=1$ \cite{GarciaGudino:2015ocw,Rojas:2024tmn}. Thus, we consider these values as a reference of their magnitudes.\\
The global strength is set by $e\, g_{V^\prime V\gamma}$. The $g_{V^\prime V\gamma}$ parameter is taken as the ratio of the $g_{V}$ and $g_{V^\prime}$ couplings, where $g_V$ and $g_V^\prime$ account for the vector interaction with the photon, in analogy to the VMD idea. We use both possibilities $g_V/g_{V^\prime}$ and $g_{V^\prime}/g_V$. They are obtained using the VMD approach \cite{Avalos:2022txl}, with $V \to l^+ l^-$ data \cite{ParticleDataGroup:2022pth}. The values of $g_V$ are given in Table \ref{tab:coup5}, for the systems under consideration.

\begin{table}[t]
\begin{center}
\renewcommand{\arraystretch}{1}
\begin{tabular}{c|c}
\hline
\hline
Coupling & Value \\
\hline
$g_{\rho^\prime}$ & $12.918$\\
$g_\rho$ & $4.962$\\
$g_{\psi(2S)}$ & $18.799$\\
$g_{J/\psi}$ & $11.219$\\
$g_{\Upsilon(2S)}$ & $61.610$\\
$g_{\Upsilon(1S)}$ & $40.435$\\
\hline
\hline
\end{tabular}
\caption{Photon-Vector meson couplings, in the VMD approach, relevant for the computation of LFV hadronic decays using dim-5 effective operators.} \label{tab:coup5}
\end{center}
\end{table}

\begin{table}[t]
\begin{center}
\renewcommand{\arraystretch}{1}
\begin{tabular}{c|c}
\hline
\hline
Coupling & Value (GeV$^{-1}$)\\
\hline
$g_{\rho^\prime\pi\gamma}$ & $0.063$\\
$g_{\rho\pi\gamma}$ & $0.206$\\
$g_{\psi(2S){\eta_c}(1S)\gamma}$ & $-0.044$\\
$g_{J/\psi\,{\eta_c}(1S)\gamma}$ & $0.264$\\
$g_{J/\psi\,{\eta_c}(2S)\gamma}$ & $-0.007$\\
$g_{\Upsilon(2S){\eta_b}(1S)\gamma}$ & $0.003$\\
$g_{\Upsilon(1S){\eta_b}(1S)\gamma}$ & $-0.043$\\
$g_{\Upsilon(1S){\eta_b}(2S)\gamma}$ & $0.002$ \\
\hline
\hline
\end{tabular}
\caption{Relevant $g_{V\!P\gamma}$ couplings for the computation of LFV hadronic decays using dim-7 effective operators. The light vector mesons couplings are obtained from VMD relations \cite{Avalos:2022txl} and the ones for quarkonia are taken from \cite{Li:2018uif, Ridwan:2024ngc}.} \label{tab:coup7}
\end{center}
\end{table}

\subsection{Dim-7 operator driven processes}

We now consider scenarios where the dim-5 operators are suppressed, while the dim-7 operators are not (as discussed in Section IV of Ref.~\cite{Fortuna:2022sxt}). We proceed by describing the vector meson decay first, followed by the Pseudoscalar one.

\subsubsection*{Vector meson decay}

The $V^\prime\to V\mu  e$ decay is a one-loop level process when driven by the dim-7 operators shown in Eq. (\ref{eq:dim7}), and the corresponding Feynman diagram is depicted in Fig. \ref{fig:dim7}. The loop involves an intermediate pseudoscalar state and a generic vector-pseudoscalar-photon interaction ($VP\gamma$), which is described by the effective Lagrangian
\begin{equation} \label{eq:VPg}
    \mathcal{L}=g_{V\!P\gamma}\epsilon_{\alpha\beta\mu\nu}\partial^\alpha V^\beta\partial^\mu A^\nu P\,,
\end{equation}
where $V$, $A$ and $P$ are the vector, photon and pseudo-scalar fields, respectively. 
The $g_{V\!P\gamma}$ couplings can be obtained from  phenomenological descriptions based on VMD for the  $\rho (1450)$ and $\rho (770)$. The value of the corresponding couplings are extracted from VMD relations between radiative and hadronic couplings, $g_{\rho^\prime\pi\gamma} = g_{\rho^\prime\omega\pi}e/g_\omega$ and $g_{\rho\pi\gamma}=g_{\rho\omega\pi}e/g_\omega$. These values were obtained from a previous analysis of low energy couplings in the same approach \cite{Avalos:2022txl}. 

The $g_{V\!P\gamma}$ couplings for quarkonia can be computed within quark models exploiting the particular features of the strongly interacting system \cite{Ridwan:2024ngc,Li:2018uif,Godfrey:1985xj, Iachello:1992yu}. The values of the $g_{V\!P\gamma}$ couplings we use in this work are shown in Table \ref{tab:coup7}.  

\begin{figure}[t]
\includegraphics[width=8cm]{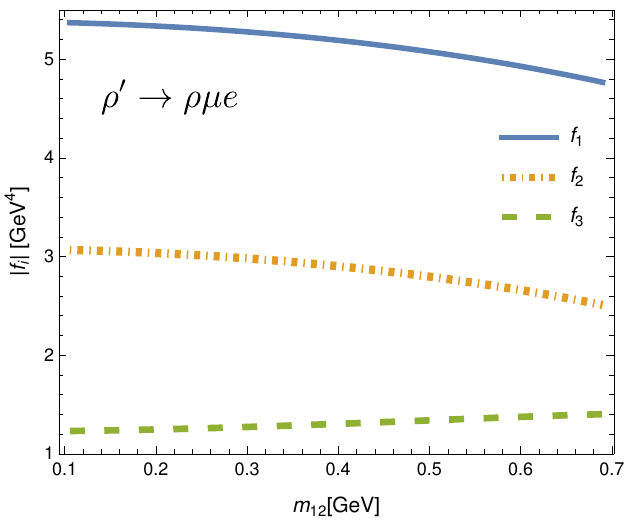}
\caption{\label{fig:loop_functions} Magnitude of the $f_1, f_2$ and $f_3$ loop functions, computed for the $\rho^\prime \to \rho \mu e$ decay as a function of the dilepton invariant mass. $|f_1|$ and $|f_3|$ are multiplied by $p \cdot q$ to be dimensionally consistent with $|f_2|$.} 
\end{figure}

The amplitude corresponding to the diagram in Fig. \ref{fig:dim7} is twofold, depending on which structure is used to describe the two photons coupled to the LFV pair. That is, proportional to the EM tensor (here labeled by $F$)  or to the dual EM tensor ($\tilde{F}$) as shown below:
\begin{align} 
    \mathcal{M}_{\text{dim7}(F)}&=2\ell^{F} \Gamma^{F}_{\alpha\beta}\eta^\alpha \epsilon^{*\beta}, \label{eq:amp_dim7}\\
   \mathcal{M}_{\text{dim7}(\tilde{F})}&=2\ell^{\tilde{F}} \Gamma^{\tilde{F}}_{\alpha\beta} \eta^\alpha \epsilon^{*\beta}, \label{eq:amp_dim7_tilde}
\end{align} 
where $\ell^{\ptwiddle{F}}=\bar{u}_1(\ptwiddle{G}^{\mu e}_{SR} P_R+\ptwiddle{G}^{\mu e}_{SL}P_L) v_2$, according to Eq. (\ref{eq:dim7}) and the structures resulting from the evaluation of the loop are $\Gamma_{\alpha\beta}^F$ and $\Gamma_{\alpha\beta}^{\tilde{F}}$, given by 
\begin{align}
    \Gamma^{F}_{\alpha\beta}&=\frac{ig_{V^\prime P\gamma}\,g_{VP\gamma}}{16\pi^2}\left\{f_1(m_{12}^2) p_\alpha q_\beta+f_2(m_{12}^2) g_{\alpha\beta}) \right\}\nonumber\,,\\
    \Gamma^{\tilde{F}}_{\alpha\beta}&=\frac{ig_{V^\prime P\gamma}\,g_{VP\gamma}}{16\pi^2} \epsilon_{\alpha\beta\mu\nu} p^\mu q^\nu f_3(m_{12}^2)\,,
\end{align}
where $f_1,f_2$ and $f_3$ are loop functions that depend on the dilepton invariant mass, $m^2_{12}\equiv (l_1+l_2)^2$. In Fig. \ref{fig:loop_functions}, we show the behavior of the magnitude of these functions computed for the $\rho^\prime \to \rho \mu e$ decay. $|f_1|$ and $|f_3|$ are multiplied by $p \cdot q$ to be dimensionally consistent with $|f_2|$. Similar features are exhibited for the other systems. 

\subsubsection*{Pseudoscalar meson decay}

\begin{figure}[t]
\includegraphics[width=8cm]{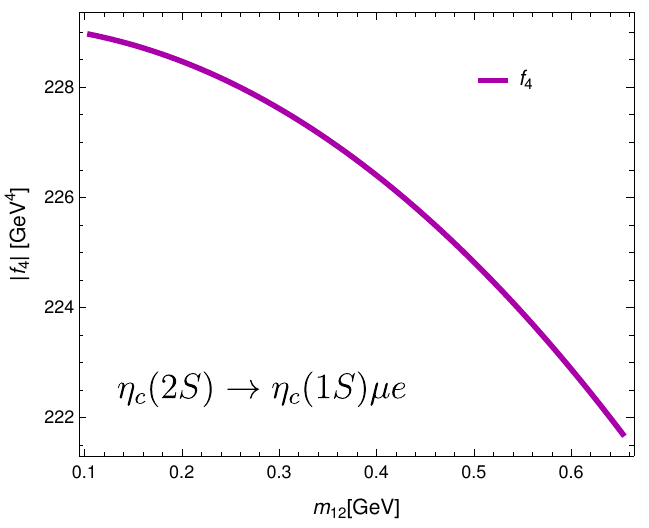}
\caption{\label{fig:loop_ps} Magnitude of the $f_4$ loop function, computed for the $\eta_c(2S)\to\eta_c(1S)\mu e$ decay, as a function of the dilepton invariant mass.} 
\end{figure}

The $P^\prime(q)\to P(p)\mu(l_1) e(l_2)$ decay (in parenthesis are the corresponding momenta) is also a one-loop level process when driven by the dim-7 operators, the corresponding Feynman diagram is depicted in Fig. \ref{fig:dim7_p}. The loop involves a generic $V\!P\gamma$ interaction, described by the effective Lagrangian in Eq. (\ref{eq:VPg}) and the couplings are taken in a similar way as in the  vector meson decays.
This decay process only receives contribution from the dim-7 operators with the EM tensor, while the operators with the dual EM tensors give a null contribution. The latter can be seen from momentum conservation and the number of degrees of freedom involved to describe the amplitude, which includes a Levi-Civita tensor. The only amplitude for this process can be written as:
\begin{equation}
    \mathcal{M}=\frac{i g_{V P^\prime\gamma}\, g_{V P\gamma}}{4\pi^2} \bar{u}_1(G^{\mu e}_{SR} P_R+G^{\mu e}_{SL}P_L) v_2 \,f_4(m_{12}^2)\,,
\end{equation}
where $f_4$ is a loop function depending on the dilepton invariant mass. In Fig. \ref{fig:loop_ps} we show the behavior of the magnitude of $f_4$ for the case of $\eta_c(2S) \to \eta_c(1S) \mu e$.
We have not considered dim-5 operators involving the $P^\prime P\gamma$ vertex, since it would require a higher order hadronic contribution associated to an effective anomalous magnetic term of the pseudoscalars.
\\

We use Package-X~\cite{Patel:2015tea} for the analytical evaluation of the loop integrals. For the numerical evaluation, we use the COLLIER library~\cite{Denner:2016kdg}, considering only the finite part and a cut-off of $\mu=1$ GeV, expected to be valid for hadronic functions. Energy dependent couplings (form factors) in the vertices would soften the divergent part, but small effects are expected given the relatively low energy transfer.

 \begin{figure}[t]
\centering
\includegraphics[width=7.5cm]{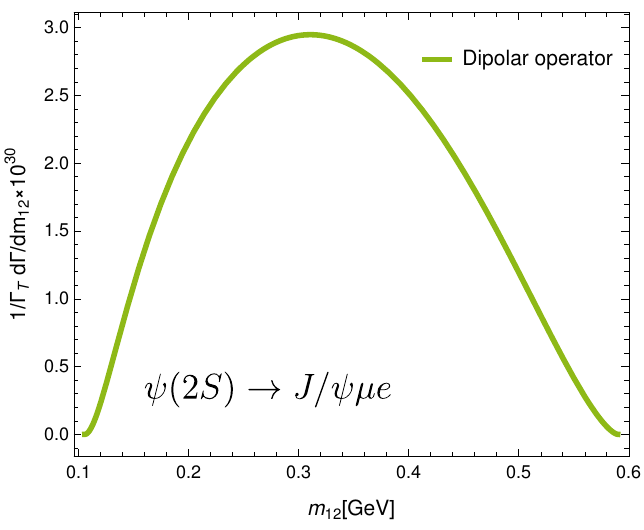}
\caption{\label{fig:dlepton5} Dilepton invariant mass distribution for the \mbox{$\psi(2S)\to J/\psi\mu e$} decay driven by the dim-5 operators, taking \mbox{$g_{\psi(2S) J/\psi \gamma}=1$, as a representative value} (cf. Eq. (\ref{eq:amp_dim5})).}
\end{figure}

\begin{figure}[t]
\centering
\subfigure{\includegraphics[width=7.5cm]{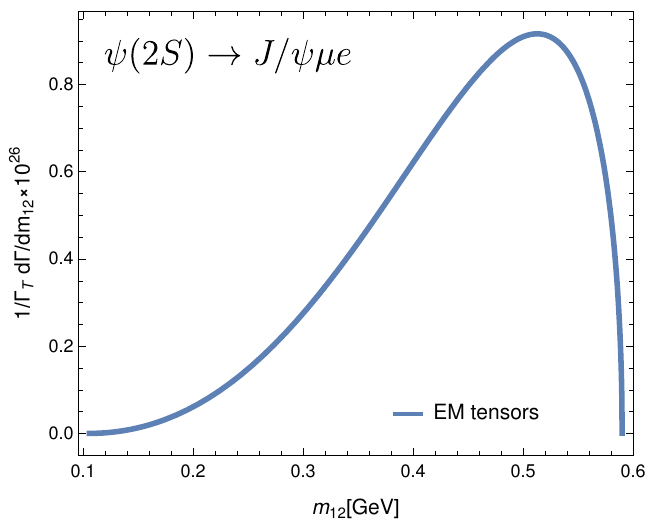}}
\subfigure{\includegraphics[width=7.5cm]{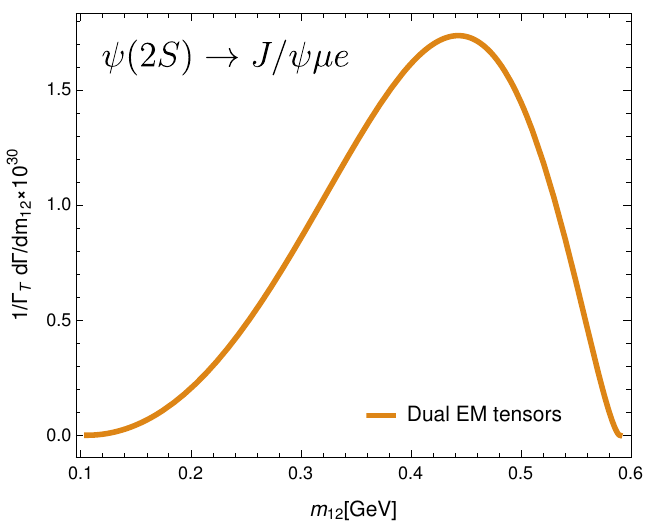}}
\caption{\label{fig:dlepton7} Dilepton invariant mass distribution for the $\psi(2S)\to J/\psi \mu e$ decay driven by the dim-7 operators. The upper panel corresponds to the EM tensor (Eq. (\ref{eq:amp_dim7})) and the lower panel corresponds to the dual EM tensor (Eq. (\ref{eq:amp_dim7_tilde})).}
\end{figure}

 \begin{figure}[t]
\centering
\includegraphics[width=8cm]{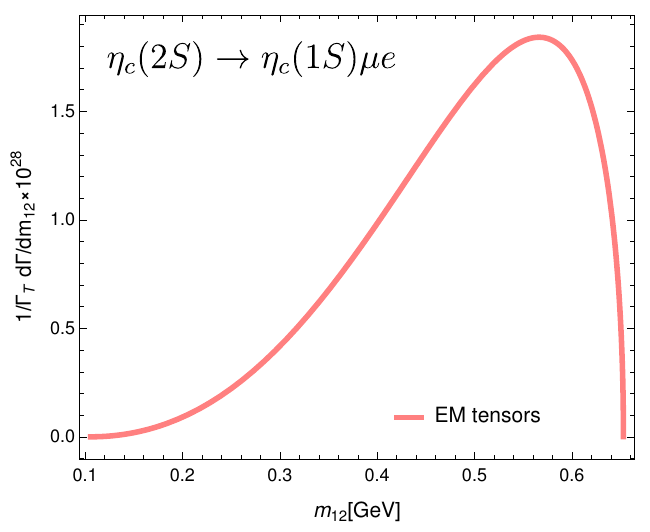}
\caption{\label{fig:ps_etac} Dilepton invariant mass distribution for the \mbox{$\eta_c(2S)\to\eta_c(1S) \mu e$} decay driven by the dim-7 operators. For this type of processes, the operators with the dual EM tensors give a null contribution.}
\end{figure}

 \section{Results and Discussion}
We compute the dilepton invariant mass distribution and the BR's for the different processes discussed previously, using the standard kinematics as given in PDG \cite{ParticleDataGroup:2022pth}.

For vector meson decays, we compute the dilepton invariant mass distribution $(1/\Gamma_T)d\Gamma/dm_{12}$, normalized to the total decay width ($\Gamma_T$). They are shown in Fig.~\ref{fig:dlepton5} for the dim-5 operators and  Fig.~\ref{fig:dlepton7} for the dim-7 operators. For the latter, the upper panel corresponds to the EM tensor (Eq. (\ref{eq:amp_dim7})) and the lower panel corresponds to the dual EM tensor (Eq. (\ref{eq:amp_dim7_tilde})). This observable allows one to distinguish the different energy behavior between them, the dim-5 operators generate a curve slightly tilted towards the left, while the dim-7 operators produce curves tilted to the right. The relative magnitude among the dim-7 operators comes from the evaluation of the loop. For example, for the $\rho^{\prime}\to\rho\mu e $, at the amplitude level the EM tensor part is an order of magnitude larger than the dual EM tensor. This is related to the magnitude of the loop functions. For quarkonia this relative difference is larger for the same reason.
Besides the relative magnitude, the $(1/\Gamma_T) d\Gamma/dm_{12}$ from the EM tensor (upper panel in Fig. \ref{fig:dlepton7}) generates a curve whose peak is more shifted towards the end of the plot with respect to the dual EM one (lower panel in Fig. \ref{fig:dlepton7}). 
These observations apply for all the vector systems under consideration. In the pseudoscalar case, since the dual EM tensor contribution is null, the comparison is only made between the EM tensor part with respect to the vector case, where we observe that they exhibit a similar behavior.

\begin{table}[t]
\begin{center}
\renewcommand{\arraystretch}{1.4}
\begin{tabular}{cc}
\hline
\hline
Operator & BR($\Upsilon(2S)\to \Upsilon(1S) \mu e$) \\
\hline
Dim-5 dipolar & $[3.9-20.9]\times 10^{-31}$\\
Dim-7 EM  & $1.6\times 10^{-26}$\\
Dim-7 Dual EM & $5.0\times 10^{-31}$\\
\hline
\hline
Operator & BR($\psi(2S)\to J/\psi \mu e$) \\
\hline
Dim-5 dipolar & $[3.2-25.0]\times 10^{-31}$\\
Dim-7 EM  & $2.1\times 10^{-27}$\\
Dim-7 Dual EM & $4.1\times 10^{-31}$\\
\hline
\hline
Operator & BR($\rho^{\prime}\to\rho\mu e$) \\
\hline
Dim-5 dipolar & $[1.7-77.8]\times 10^{-33}$\\
Dim-7 EM  & $1.7\times 10^{-32}$\\
Dim-7 Dual EM & $4.4\times 10^{-34}$\\
\hline
\hline
\end{tabular}
\caption{Branching ratios involving de-excitations of the $\rho^\prime$ meson and quarkonia, driven by the dim-5 and dim-7 operators. The interval obtained by using the dipolar operator corresponds to the values we used for the $g_{V^\prime V\gamma}$ coupling.} \label{tab:VBRsQ} 
\end{center}
\end{table}

\begin{table}[t]
\begin{center}
\renewcommand{\arraystretch}{1.4}
\begin{tabular}{cc}
\hline
\hline
Operator & BR($\eta_b(2S)\to\eta_b  (1S) \mu e$) \\
\hline
Dim-7 EM  & $7.7\times 10^{-29}$\\
\hline
\hline
Operator & BR($\eta_c(2S)\to\eta_c(1S) \mu e$) \\
\hline
Dim-7 EM  & $4.7\times 10^{-29}$\\
\hline
\hline
\end{tabular}
\caption{Branching ratios involving de-excitations of Pseudoscalar states, driven by the dim-7 operator.} \label{tab:PBRsQ} 
\end{center}
\end{table}

The BR's for the $\rho^\prime$ meson and quarkonia, obtained using the dim-5 and dim-7 operators are shown in Table \ref{tab:VBRsQ}. For the dim-5 operators we quote an interval for the BR's, obtained from the different values used to compute the $g_{V^\prime V\gamma}$ couplings, from the ratio of the corresponding $g_V$ and $g_{V^\prime}$ parameters. That is, they produce a difference of around two orders of magnitude for the $\rho^\prime$, while for quarkonia the difference is only of one order of magnitude. We also explored the structure of the $V^\prime V\gamma$ vertex, Eq. (\ref{eq:dip-cuad}), and found that the dominant contribution comes from the dipolar magnetic interaction, while the quadrupolar electric interaction has a relative suppression of about two or three orders of magnitude (depending on the specific process).

The difference in magnitude for the BR's of the two dim-7 contributions follows the same argument as in the dilepton invariant mass distribution. It is interesting to note that the relative magnitude is opposite to the one obtained from interactions in nuclei \cite{Fortuna:2023paj} where the dominant contribution is driven by the dual EM tensor, while in our case it is the EM tensor (see Table \ref{tab:VBRsQ}). This feature offers the possibility to disentangle both contributions from complementary observables. However, notice that this is not a general statement, since the observation is drawn from two different systems: while we are working with an effective vertex involving an electron and a muon, in Ref.~\cite{Fortuna:2023paj} they analyzed  the light lepton to tau conversion.

The less suppressed BR contribution comes from  dim-7 EM term for the $\Upsilon(2S)\to \Upsilon(1S) \, \mu e$ decay, with a BR of $\mathcal{O}(10^{-26})$, followed by the same operator in the $\psi(2S)\to J/\psi \, \mu e$ decay, with a BR of $\mathcal{O}(10^{-27})$.
To give an idea of the meaning of such magnitudes, we can compare them with the case of vector mesons LFV annihilation, Table \ref{tab:others}, which exhibit similar orders of magnitude. That is, the de-excitation processes may be as relevant/restricted as the direct annihilation ones.
The BR's for the Pseudoscalar states, obtained using dim-7 operators, are shown in Table \ref{tab:PBRsQ}. We have considered both bottomonium and charmonium states, and found that they have similar branching ratios of $\mathcal{O}(10^{-29})$, regardless of the quark content.

In this work, we have explored the hadronic de-excitation via LFV, a scenario that has not been considered in the literature, and the different features they can exhibit, depending on the effective operator producing the LFV pair.  
We have considered both vector and pseudoscalar excited states, involving light and heavy quarks systems. 
Our results for the BR's are comparable with the BR's from direct LFV decays of quarkonia \cite{Calibbi:2022ddo}. The magnitude of these indirect bounds strongly relies on the constraints from direct experimental searches. Thus, complementary information can be extracted between them.
The dilepton invariant mass distributions show that they may be useful to impose constraints on individual contributions and help to disentangle them, when complemented with observables from nuclei.
Our results are obtained in an EFT approach, corresponding to general estimates. Specific models can provide results (naturally model dependent) which are less restrictive, taking into account specific matter content and interactions.

Current experiments such as LHCb, Belle II and BES III are reaching a high luminosity stage, where these kind of hadrons are copiously produced, as we already mentioned before. They might provide the opportunity to explore this type of scenario not yet considered, and place upper bounds on them. 
Our predictions rely on the mechanism driving the $\mu \to e \gamma$. Therefore, finding any evidence of LFV de-excitation at a larger BR, would imply the existence of a new LFV mechanism not captured by our effective approach. The case of the $\rho^\prime$ is interesting as it illustrates the mechanism with well known hadrons, but the experimental limitations and its large decay width makes it a very unlikely possibility. On the other hand, quarkonia is a more favorable case due to their small decay widths.
The exploration of non-conventional processes, as the one presented here, contributes to get a better understanding of the link between LFV and hadronic de-excitation processes.\\

\begin{acknowledgments}
We thank Pablo Roig and Roelof Bijker for very useful observations. The work of F.F. is funded by \textit{Estancias Posdoctorales por México, Estancia Posdoctoral Iniciales, CONAHCYT-SECIHTI}. We acknowledge the support of CONAHCYT-SECIHTI, México, and the support of DGAPA-PAPIIT UNAM, under Grant No. IN110622.
\end{acknowledgments}

\vspace{3mm}
Data Availability Statement: No Data associated in the manuscript.    

\bibliography{rholfv}

\end{document}